\documentclass[aps,epsf,twocolumn,epsfig,showpacs]{revtex4-1}
\usepackage{graphics}
\usepackage{graphicx}
\usepackage{dcolumn} 
\usepackage{bm}
\usepackage{epsfig}
\usepackage{float}
\begin{document}
\title{Tuning the Eu valence in EuPd$_3$B$_x$: pressure versus valence electron count - \\ a combined computational and experimental study}
\author{M. Schmitt$^1$, R. Gumeniuk$^1$, A. Trapananti$^{2,3}$, G. Aquilanti$^{4,3}$, 
C. Strohm$^3$, K.~Meier$^1$, U. Schwarz$^1$, M. Hanfland$^3$, W. Schnelle$^1$, A. Leithe-Jasper$^1$, H. Rosner$^1$}
\affiliation{$^1$Max-Planck-Institut f\"ur Chemische Physik fester
Stoffe, 01187 Dresden, Germany}
\affiliation{$^2$ CNR-Instituto Officina dei Materiali-OGG, c/o ESRF, BP 220 38043 Grenoble Cedex 9, France}
\affiliation{$^3$European Synchrotron Radiation Facility, BP 220, 38043 Grenoble Cedex 9, France}
\affiliation{$^4$ Sincrotrone Triest, SS 14 km 16.5 in Area Science Park, Basovizza, Trieste, 34149 Italy}

\date{\today}
\begin{abstract}

In a joint theoretical and experimental study we investigate the pressure dependence of the Eu valence in EuPd$_3$B$_x$ ($0 \le x \le 1$). Density functional band structure calculations are combined with x-ray absorption and x-ray diffraction measurements under hydrostatic pressures up to 30 GPa. It is observed that the heterogenous mixed-valence state of Eu in  EuPd$_3$B$_x$ ($x\ge 0.2$) can be suppressed partially in this pressure range. From the complementary measurements we conclude that the valence change in  EuPd$_3$B$_x$  is mainly driven by the number of additional valence electrons due to the insertion of boron, whereas the volume change is a secondary effect. A similar valence change of Eu in Eu$_{1-x}$La$_x$Pd$_3$ is predicted for $x \ge 0.4$, in line with the suggested electron count scenario.

\end{abstract}

\maketitle

%

\section{Introduction}

The possibility to observe and induce valence instabilities in Eu containing intermetallic compounds has been of considerable interest and the topic of many experimental as well as theoretical investigations \cite{Bauminger1974, Felner1977,Sampathkumaran1981, Perscheid1985, Nagarajan1981, Wortmann1991, Wada1999, Ni2001, Hossain2004, Michels1995, Poettgen2000, Klauss1997, Ksenofontov2006}.
In fact, europium together with ytterbium are the only lanthanide metals which are divalent (configurations 4$f^7$; 4$f^{14}$) in their elemental metallic standard state, as well as in some alloys and intermetallic compounds \cite{Gschneidner1969}. In that case, their crystal chemistry strongly resembles that of the alkaline-earth metals, typically influenced by the large radii of the ions. Nevertheless, in many systems Eu is found to be in a trivalent (4$f^6$) state, similar to the majority of the rare-earth metals \cite{Buschow1979}. Accordingly, the physical properties of Eu are significantly different in the two states, namely Eu$^{2+}$ (S = 7/2, L = 0) carrying a high magnetic moment (J = 7/2) compared to Eu$^{3+}$ with a non-magnetic ground state (S = 3, L = 3, J = L-S = 0) and low-lying excited magnetic states (J = 1, 2,...). Many factors like the local environment (determining the crystal field splitting), the electronegativity and the concentration of alloying partners \cite{M76,B79,M83} as well as external parameters like temperature, pressure and magnetic field \cite{L81} determine and influence the valency of Eu in a compound. 

In this context it has been known and controversially discussed for a considerable time that insertion of boron in cubic EuPd$_3$ (Cu$_3$Au type of structure) induces a change of the  Eu valence from an essentially non-magnetic Eu$^{3+}$ ($4f^6$) state into a strongly magnetic Eu$^{2+}$ ($4f^7$) state (see Ref.~\onlinecite{gumeniuk10}).   For boron contents passing a threshold value $x_c$ in the system EuPd$_3$B$_x$, a heterogeneous mixed-valence state was inferred to exist, including the stoichiometric compound EuPd$_3$B.  However, a recent study of the electronic structure of intermetallic $RE$Pd$_3$B$_x$ borides covering the whole series from $RE =$ La to Lu as a function of the boron content \cite{CL07} called the stability range of EuPd$_3$B$_x$ into question. In order to clarify this issue, we reinvestigated in a joint theoretical and experimental study a large series of EuPd$_3$B$_x$ and GdPd$_3$B$_x$ compounds \cite{gumeniuk10}.  By x-ray diffraction (XRD), metallography, energy and wave-length dispersive x-ray spectroscopy, as well as chemical analysis, 
the homogeneity ranges for EuPd$_3$B$_x$ and  GdPd$_3$B$_x$ could be established as $x \le 0.53$ and $x \le 0.42$, respectively.
Density functional (DFT) based electronic structure calculations predicted a valence change in EuPd$_3$B$_x$ above $x_c = 0.19(2)$ from a non-magnetic Eu$^{3+}$ ground state to a magnetic Eu$^{2+}$ state, which is reflected in a discontinuity of the lattice parameter. In contrast, the GdPd$_3$B$_x$ alloy system with a stable Gd$^{3+}$ state exhibits an almost linear increase of the lattice parameter following Vegard's law. Consistent with the theoretical calculations, the lattice parameter vs. $x$ indeed shows a kink for EuPd$_3$B$_x$ at $x_c = 0.22 (2)$. X-ray absorption spectroscopy (XAS) in line with magnetic susceptibility and specific heat data assign this kink to a transition into a heterogeneous mixed valence state for Eu associated with a change of the mean Eu valence from Eu$^{3+}$ ($x \le 0.2$) towards Eu$^{2.5+}$ ($x = 0.5$).

This close interplay  of the Eu valence state and the discontinuity in the unit cell volume under boron insertion raises the question, if the valence transition is driven by the  mere change of interatomic distances, by the B chemistry, or ruled by the valence electron count. To clarify this issue, we investigate the influence of high pressure on EuPd$_3$B$_x$ compounds with fixed B content above the valence transition ($x\ge x_c$).  
In our electronic structure calculations, we find a subtle interplay between doping and volume effects  on the valence state, while the disorder within the system is less relevant. However, being aware that present-day DFT calculations have still difficulties with respect to the description of strong correlation and disorder, we challenge our theoretical results by a combination of XAS and XRD experiments under high pressure. These measurements allow to follow the questions, (i) if we can observe the qualitatively predicted change of Eu valence under pressure for a fixed B content experimentally and (ii) if the transition is continuos as under B insertion. Finally, we elucidate the driving force for the onset of the transition. Our results show, that applying pressure can reverse the effect only partially, driving the magnetic Eu$^{2+}$ state towards the non-magnetic Eu$^{ 3+}$ state and thus point to an important contribution of the B chemistry or the electron count to the valence transition, ruling interatomic distance effects out.  
Recently, a similar valence scenario 
was observed for Eu$_{0.4}$La$_{0.6}$Pd$_3$ \cite{la_pap}.
Applying our calculational approach, we find a stable Eu$^{2+}$ state in  Eu$_{0.4}$La$_{0.6}$Pd$_3$ due to the La insertion, in agreement with the reported experiments. A systematic, theoretical investigation of the whole series of Eu$_{1-x}$La$_x$Pd$_3$ ($0\le x \le1$) compounds allows to predict a critical La concentration for the transition
 and supports independently the predominant role of charge doping for the Eu valence change.

The manuscript is organized as follows. All technical details concerning our experimental and theoretical methods are 
shortly summarized in the next section. Then, the results of our theoretical approach (fixed-spin moment calculations) are presented (Sec.~\ref{res_calcs}) followed by high-pressure XAS (Sec.~\ref{res_xas}) and XRD (Sec.~\ref{res_xrd}) measurements to study the stability of the valence state under volume change for EuPd$_3$B$_x$.
Finally the influence of an alternative La substitution is investigated (Sec.~\ref{res_lasub}) by our computational approach allowing to distinguish between the
effects caused by B chemistry or pure valence electron change. A summary and outlook concludes the manuscript (Sec.~\ref{sum}).

\section{Methods}\label{method}

To study the influence of changes in volume and in number of valence electrons (induced by the B insertion) on the Eu valence state, electronic structure calculations have been performed using 
the full potential non-orthogonal local-orbital minimum basis scheme FPLO (version: fplo5.00-19) within the local spin density approximation (LSDA).\cite{fplo1}
Within the scalar relativistic calculations the exchange and correlation potential of Perdew and Wang was chosen \cite{PW92}.
The same basis set as in Ref.~\cite{gumeniuk10} was used, treating the rare-earth $4f$ states as valence states.
The strong correlation of Eu 4$f$ electrons was considered in a mean-field way by the LSDA+$U$ approximation \cite{fplo2} 
applying an on-site Coulomb repulsion $U=8$\,eV and on-site exchange $J=1$\,eV and using the atomic limit for the double counting term. The variation of $U$ from 6 to 8\,eV and $J$ from 0 to 1\,eV does not change the results qualitatively. 

The disordered insertion of B for different concentrations between EuPd$_3$ and (hypothetical) EuPd$_3$B and the gradual, disordered substitution 
of La from EuPd$_3$ to LaPd$_3$ was simulated by the coherent potential approximation (CPA)\cite{koep97}.  
To check the accuracy of the LSDA+$U$+CPA calculations \cite{BEB} we compared them with the results for the ordered structures ($x$=0, $x$=0.5 and $x$=1).
Furthermore, to study the role of electron doping independent from the rare-earth site and from structural changes, for EuPd$_3$ the virtual crystal approximation (VCA)  at the Pd site was used.

Polycrystalline samples were prepared by arc melting of the elements under Ar atmosphere and carefully characterized (for details see Ref.~\onlinecite{gumeniuk10}). 
 X-ray diffraction measurements (XRD) for pressures up to 30\,GPa were performed at the high-pressure beam-line ID09 of the ESRF at room temperature for two different EuPd$_3$B$_x$ samples ($x=0.32, 0.48$) above the valence transition and two GdPd$_3$B$_x$ compounds with corresponding B content as reference system. For best possible hydrostatic conditions we used a membrane diamond anvil cell (DAC) with helium as 
 pressure medium. The pressure was determined using the ruby fluorescence method \cite{forman72}.
 The collected patterns were integrated using the program Fit2D \cite{fit2D}. After a background correction the data were refined with the fullprof package \cite{fullprof}.
 Since  rhombohedral and orthorhombic distortions frequently occur in compounds with perovskite- and related  structures \cite{Howard2004, Aleksandrov1976}, all XRD peaks in the measured powder patterns were carefully analyzed with respect to  possible splitting. In the pressure range of 15-30 GPa full-width of half maxima (FWHM) of the reflections increase slightly with increasing diffraction angle without any signs of splitting. On the other hand, in the low pressure range (0-15 GPa) some peaks in a few XRD patterns were broadened, possibly indicating a splitting (see Ref.~\onlinecite{note4}). However, the observations are not  systematic and do not obey any rhombohedral, tetragonal or orthorhombic rules. Therefore we can conclude that these observations are most likely an artifact of the measurement.

Eu $L_{\mathrm{III}}$-edge (6977 eV) x-ray absorption spectroscopy (XAS) measurements were performed at the energy dispersive XAS beam-line ID24 of the ESRF, on two EuPd$_3$B$_x$ samples ($x= $0.32, 0.48).
The powder samples were pressurized up to 25\,GPa within a non-magnetic Cu-Be DAC using silicon oil as pressure transmitting medium. As in the high-pressure XRD experiment, a ruby chip was used for the pressure determination. The obtained XAS spectra were analyzed by a least-squares fitting procedure to determine the average Eu valence $\nu$ by the relative intensities of Eu$^{2+}$ and Eu$^{3+}$ components, as described in Ref.~\onlinecite{MizTsu2007}.

\begin{figure*}
\begin{center}\includegraphics[
  width=15cm,
  angle=0]{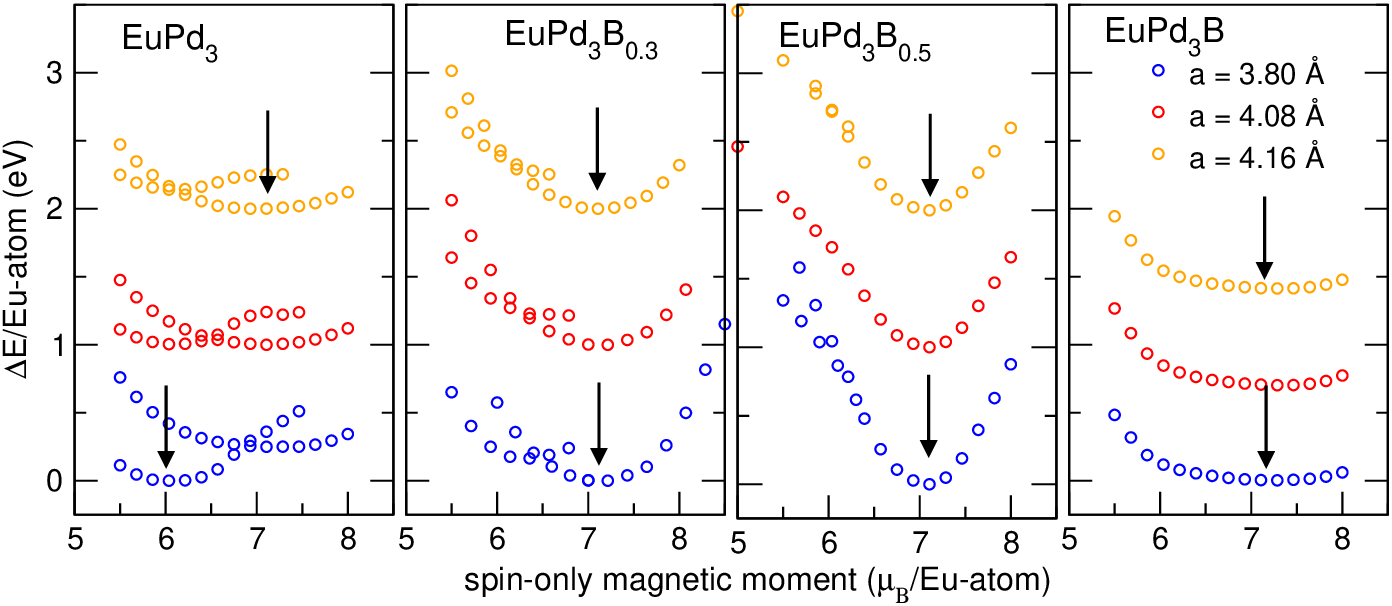}\end{center}
\caption{\label{fsm}(color online)
Fixed spin moment calculations for four different boron contents $x$ in EuPd$_3$B$_x$ under pressure (for different volumes).
The total energies for the two ordered structures EuPd$_3$ and EuPd$_3$B are calculated by LSDA+$U$. 
For the compounds with $x=0.3$ and $x=0.5$ the CPA approach was used simulating a disordered B insertion.
The total energies are given relative to the corresponding minima. Arrows indicate the global minimum for a specific B content and volume. 
}\end{figure*}
 
\section{Results}
\subsection{Calculation}\label{res_calcs}

To elucidate the origin of the valence transition in EuPd$_3$B$_x$ we try to separate the effects caused by a change of the unit cell volume from the effects of B substitution. 
 Unfortunately it is not possible to simulate directly the mixed valence state observed in the experiments, as the LSDA+$U$ method favors full polarization and therefore suppresses a fractional occupation of the Eu $4f$ states. However, it is possible to stabilize Eu $4f$ occupations close to the two limiting cases of 6$\mu_B$/Eu and 7$\mu_B$/Eu in the majority spin channel:  Starting from these two solutions we performed fixed-spin moment calculations for compounds with fixed B concentration and  varied the volume of the unit cell \cite{note1}. As the spin component of the moment in these calculations is related to the  Eu $4f$ occupation and therefore to the valence state of Eu, the interplay of volume and $4f$ charge effects can be studied.
   The comparison of the total energies yields the more stable configuration for a specific
  B content -- volume combination.  
 In Fig.~\ref{fsm} the dependence of the total energy on the spin-only moment for four compounds with different B content is depicted. While in the case of EuPd$_3$ the two branches of the energy curves regarding the two limiting Eu $4f$ occupations are clearly visible, the stabilization of the Eu 4$f^6$ phase is progressively suppressed with increasing B content. 
 
In more detail, we find a subtle balance between the Eu$^{3+}$ and Eu$^{2+}$ state for EuPd$_3$ in its optimized equilibrium volume ($a=4.08$\AA, see left panel of Fig.~\ref{fsm}, middle graph). The energy difference between the two solutions is less than 4\,meV/Eu-atom. 
Applying pressure, simulated by a reduction of the unit cell volume ($a=3.80$\AA),
clearly stabilizes the Eu$^{3+}$ state by an energy difference of about 0.25 eV/Eu-atom. If the unit cell volume is expanded ($a=4.16$\AA) compared to the equilibrium volume,  the Eu$^{2+}$ state becomes energetically preferred ( arrows in Fig.~\ref{fsm}). Thus, rin our calculations, the valence state of Eu in EuPd$_3$  exhibits a clear volume effect.

In the case of EuPd$_3$B$_{0.3}$ with a B content just above the onset of the valence transition \cite{gumeniuk10}, the two solutions are still well pronounced. Although their total energy shifts against each other under pressure, the decrease of the unit cell volume does not change the position of the global energy minimum. Consequently, the Eu$^{2+}$ state remains stable in the whole volume range, in contrast to the behavior of EuPd$_3$.
This stabilizing effect of the B insertion on the Eu$^{2+}$ state is strengthened further with increasing B content. While for EuPd$_3$B$_{0.5}$ the metastable solution of Eu$^{3+}$ can be obtained only for spin-only moments below $6\mu_B/$Eu with a sizable energy difference to the Eu$^{2+}$ state, EuPd$_3$B converges always to the single solution of Eu$^{2+}$ (compare right panel of Fig.~\ref{fsm}).

To ensure that the observed effects are independent of a specific approximation to model the partial occupation of the B site, the results using the
CPA, VCA  and LSDA+$U$ for ordered structures ( $x=$0, 0.5, 1) were checked and compared carefully with each other (see supplementary material \cite{supp}).
From their good agreement, we conclude that the magnetic properties of the compound depend sensitively on the B content, but not on the particular B order, in contrast to an earlier study, where the Eu valence instability was connected to an anisotropic B environment \cite{cianchi91}. 
 
Our calculations yield a clear volume effect on the Eu valence state of the EuPd$_3$B$_x$ compounds, although a change of the preferred Eu $4f$ occupation is only observed for EuPd$_3$.
Besides this volume dependence, our calculations predict a strong influence of the B substitution 
on the stability of the Eu$^{2+}$ state.
This prediction and the difficulty to describe the heterogeneous mixed-valence state of Eu by the LSDA+$U$ approach in a realistic way
require experimental support for a reliable, more quantitative picture.

\subsection{XAS under pressure}\label{res_xas}

The interplay between the occupation of the Eu 4$f$ states and the
unit cell volume suggested by our calculations proposes to tune the Eu valence state 
by pressure for a fixed B concentration or even to reverse the valence transition. 

\begin{figure}[tbh]
\begin{center}\includegraphics[%
  clip,
  width=6cm,
  angle=0]{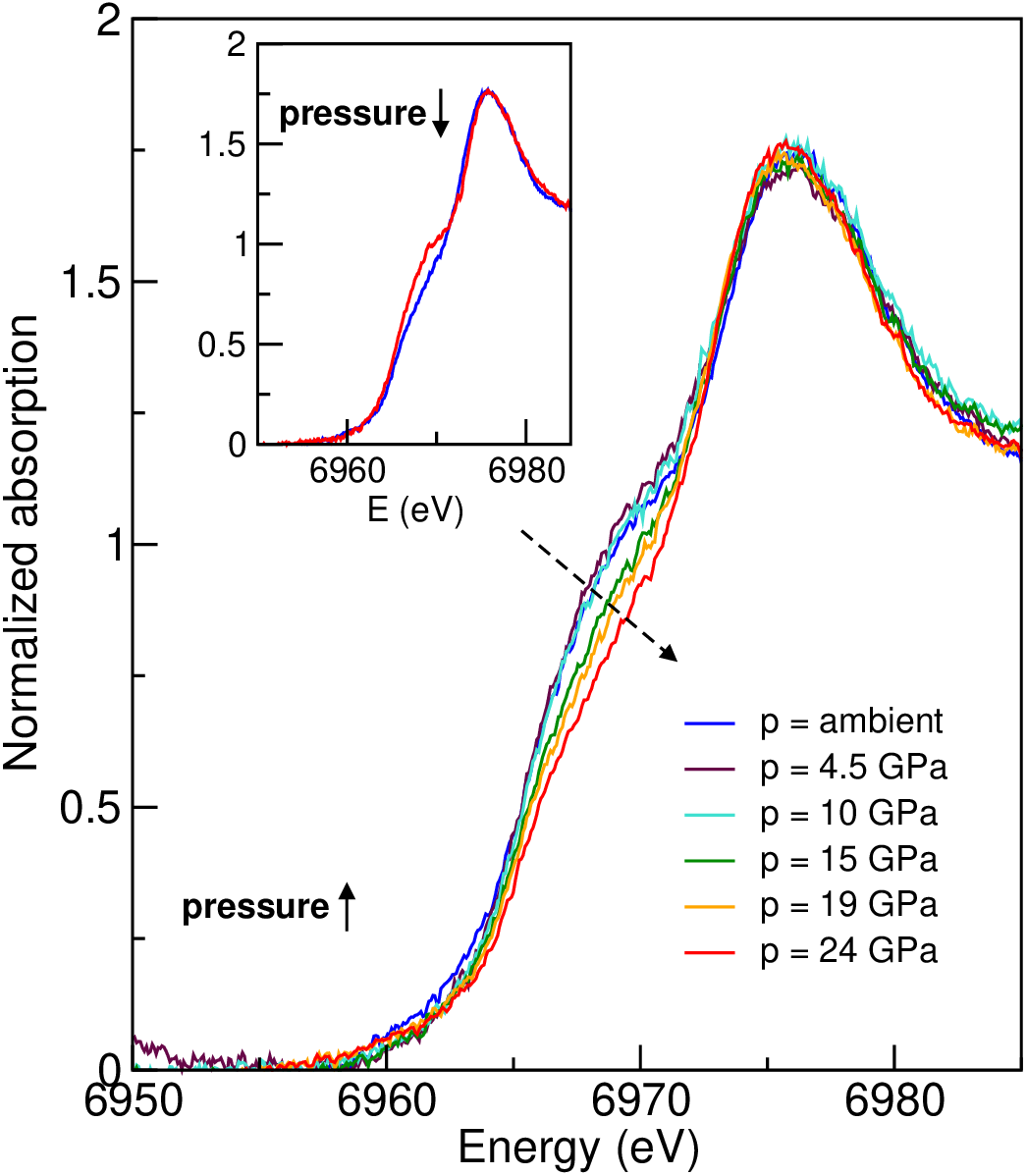}\end{center}
\caption{\label{xas}(color online)
Main panel: XAS under increasing pressure up to 25\,GPa for EuPd$_3$B$_{0.48}$. Increasing pressure 
	reduces the intensity of the $4f^7$ shoulder indicated by a dashed arrow. The effect is reversible for decreasing pressure (see inset). }
\end{figure}

As a direct probe of the Eu valence state XAS under pressure was carried out for
EuPd$_3$B$_{0.48}$ above the critical boron concentration with a mean valence $\nu=2.63(2)$. 
The obtained experimental data at room temperature are depicted in Fig.~\ref{xas}. At
ambient pressure, the measured spectrum at the $L_{\mathrm{III}}$ edge exhibits a main peak centered at 
6976.5(5)\,eV
corresponding to Eu$^{3+}$ states, and a shoulder at about  8\,eV lower in energy, originating from 
Eu$^{2+}$ states, in good agreement with earlier measurements \cite{gumeniuk10} and  the heterogeneous 
mixed -valence state of Eu. 

The stepwise increase of pressure reduces the intensity of the $4f^7$ shoulder significantly, though it is not suppressed totally.
For pressures up to 25\,GPa, the pure Eu$^{3+}$ state as present in the compounds EuPd$_3$B$_x$ with $x\le 0.2$, cannot be reached.
Thus, EuPd$_3$B$_{0.48}$ remains in a mixed-valence state but with a clearly enlarged mean valence $\nu=2.68(2)$. 
This process is reversible by decreasing pressure (compare inset of Fig.~\ref{xas}). 

For the compound EuPd$_3$B$_{0.32}$, much closer to the critical concentration for the onset of the valence transition, the two valence states should be closer
in energy according to our DFT calculations (EuPd$_3$B$_{0.3}$ exhibits a second local minimum in the energy versus moment curve
and smaller energy differences between the Eu $4f^6$ and $4f^7$ states compared to EuPd$_3$B$_{0.5}$)
 and thus more sensitive to pressure.
Unfortunately, the experimental detection of any definite valence change under pressure for such small B contents was beyond the resolution of the experimental 
set up. The contribution of Eu$^{2+}$ states to the XAS spectrum at ambient pressure is too small to allow a reliable observation of the transition as the small differences are of the same size as the background fluctuations.

Applying pressure drives the Eu$^{2+}$ states in EuPd$_3$B$_{0.48}$ towards Eu$^{3+}$ in a continuous way, 
in analogy to the continuous valence transition upon B insertion \cite{gumeniuk10}. 
However, even for pressures up to 25\,GPa the pure Eu$^{3+}$ state for EuPd$_3$B$_{0.48}$ cannot be recovered completely, raising the question whether this observation is
due to the limit of applied pressure or impeded by an intrinsic property of the compound.

\subsection{XRD under pressure}\label{res_xrd}

To estimate the pressure required to restore the equilibrium volume for the pure Eu$^{3+}$ state of the reference system EuPd$_3$, 
and to ensure the stability of the 
crystal structure under pressure, we applied x-ray diffraction (XRD) measurements upon pressure.
In addition, earlier XRD measurements allowed the indirect observation of the valence transition in 
the series of EuPd$_3$B$_x$ compounds 
by a pronounced kink in the plot of the lattice parameter {\it vs.} boron content \cite{gumeniuk10}.

\begin{figure}[htb]
\begin{center}\includegraphics[%
  clip,
  width=8cm,
  angle=0]{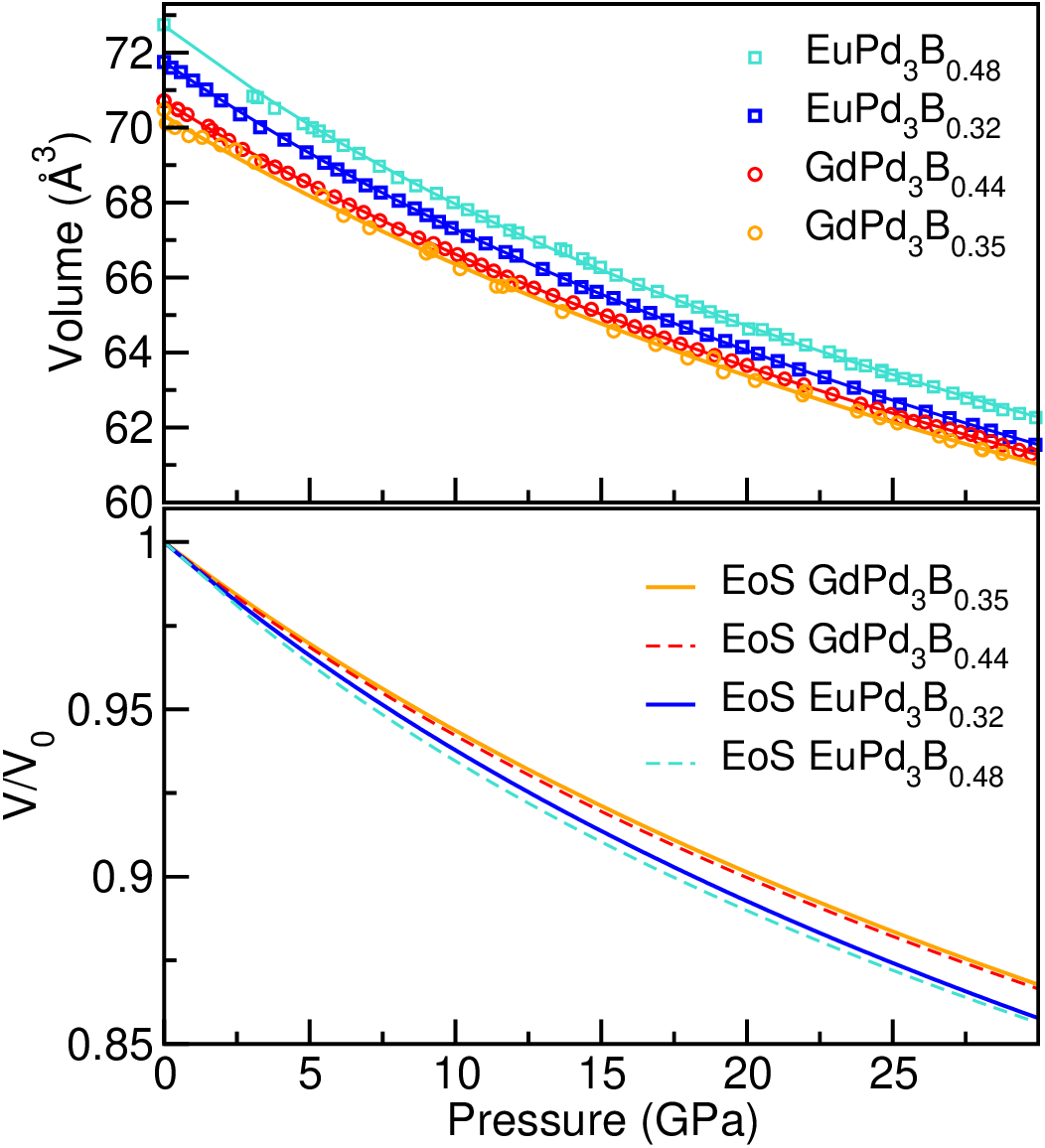}\end{center}
\caption{\label{xrd}(color online)
Top: Measured equation of states (volume vs. pressure) for two EuPd$_3$B$_{x}$ compounds above the valence transition (squares) and 
two corresponding GdPd$_3$B$_{x}$ reference systems (circles) up to 30\,GPa.  
Bottom: Comparison of the EoS, normalized to the equilibrium volume $V_0$, for all four compounds.
The data separate into two sets. The EuPd$_3$B$_{x}$ compounds 
  show clearly a smaller bulk-module compared to the Gd-reference systems, whereas the influence of differences in the B content has only minor influence.
}
\end{figure}

For pressures up to 30\,GPa no structural changes were observed for EuPd$_3$B$_{0.32}$ and EuPd$_3$B$_{0.48}$.
The measured and evaluated data are shown in Fig.~\ref{xrd} (top, open squares). To evaluate the equation of state we fitted 
the experimental volume-pressure dependence by an inverse Murnaghan-Birch equation of state  (EoS)
$$V(p)=V_0 \cdot \left(\frac{B^{'}_0 \cdot p}{B_0}+1\right)^{-1/B^{'}_0}$$
with the bulk modulus of $B_0$ and its pressure derivative $B^{'}_0$. 
The obtained EoS follows perfectly the experimental data in the whole pressure range with $B_0=133\pm 1$ ($B^{'}_0=4.7\pm 1$) for EuPd$_3$B$_{0.32}$ and $B_0=125\pm 3$ ($B^{'}_0=4.7\pm 1$) for EuPd$_3$B$_{0.48}$, respectively  (see Fig.~\ref{xrd}, upper panel and Tab.~\ref{tab1}).
Any direct structural anomaly indicating a valence instability is absent.
However, the measured EoS allows to evaluate the volume change of the compounds in the applied pressure range. For a pressure of 30\,GPa the volume for the EuPd$_3$B$_x$  compounds decreases by about 14\%, which is more than twice the 
volume difference observed under B insertion
between the boron rich EuPd$_3$B$_{0.53}$ and the boron free parent compound EuPd$_3$  \cite{note3}.
Considering the volume change as the driving force of the valence transition, a pure Eu$^{3+}$ valence state would already be expected for pressures of at most 10\,GPa, in contrast to the experimental observations.
Thus, the valence state of  EuPd$_3$B$_x$ is predominantly ruled by the inserted B, while volume changes have minor influence. 
These findings are in agreement with the trends obtained from DFT calculations, where in the case of EuPd$_3$ the volume has a sizable influence on the preferred valence state, while the insertion of B stabilizes the Eu$^{2+}$ state significantly (compare Fig.~\ref{fsm}).

Being aware of the experimental difficulty to resolve the small effects expected from a smooth and only partial change of the valence states, in combination
 with  the high-pressure method (small shear forces), we compared the results for EuPd$_3$B$_x$ with the pressure-behavior of GdPd$_3$B$_x$ ($x=0.35, 0.44$) compounds, which are used as reference systems with half filled $4f$ shell and 
therefore a stable $4f$ valence state. 
Evaluating the EoS for the GdPd$_3$B$_x$ compounds we obtain a significantly increased $B_0=152\pm 10$ 
( $B^{'}_0=4.8\pm 1$) for GdPd$_3$B$_{0.35}$  
and  $B_0=145\pm 1$ ( $B^{'}_0=4.8\pm 1$) for GdPd$_3$B$_{0.45}$, respectively (compare Tab.~\ref{tab1} and see \cite{note2}) in comparison to the respective EuPd$_3$B$_x$ systems.

For a more direct comparison, we normalized the derived EoS to their corresponding equilibrium volume (compare Fig.~\ref{xrd}, bottom). The normalized volume {\it vs.} pressure curves separate into two sets, namely the Eu and the Gd containing compounds,
pointing to a minor influence of the specific B content to the EoS.
Furthermore, this comparison demonstrates
clearly the smaller bulk moduli for the EuPd$_3$B$_x$ compounds compared to the Gd reference systems. 
 In conclusion, the "softer" pressure dependence for the EuPd$_3$B$_x$ ($x \ge 0.32$) compounds compared to the Gd reference system can be assigned to the Eu valence instability.
This trend is also found independently from a theoretical estimate of $B_0$ based on our DFT calculations, in line with the well-known problem of over-binding in LDA
the calculations result in 1\% smaller equilibrium volumes $V_0$ and about 10\% larger $B_0$, .

\begin{table}[htb]
\begin{ruledtabular}
\begin{tabular}{l cc}
 & $V_0$ & $B_0$  \\
\hline
theor. & & \\
EuPd$_3$B$_{0.3}$ & 70.92 & 139$\pm$2 \\
EuPd$_3$B$_{0.5}$ & 72.17 & 145$\pm$1\\
exptl. & &  \\
EuPd$_3$B$_{0.32}$ & 71.73 & 133$\pm$1  \\
EuPd$_3$B$_{0.48}$ & 72.72 & 125$\pm$3  \\
\hline 
theor. & & \\
GdPd$_3$B$_{0.35}$ &  69.2& 168$\pm$1  \\
GdPd$_3$B$_{0.45}$ & 70.45 & 168$\pm$1 \\
exptl. & &  \\
GdPd$_3$B$_{0.35}$ & 70.3 & 152$\pm$10 \\
GdPd$_3$B$_{0.44}$ & 72.72 & 145$\pm$1 \\
\end{tabular}
\end{ruledtabular}
\caption{\label{tab1}Fit of equations of state to the experimental data (exptl.) and the results of 
 band structure calculations for comparison (theor.).}
\end{table}

The predominant role of B insertion in the valence transition (compared to a mere volume effect) 
suggests two possible mechanisms influencing the system:  (i) 
the predominantly covalent character of the boron chemical bonding, or
(ii) an increased number of valence electrons. Stimulated  by a recent report \cite{la_pap}, we try to separate these two effects by a theoretical study of  the related system Eu$_{1-x}$La$_x$Pd$_3$, which allows to increase the number of additional valence electrons without the insertion of B. 

\subsection{La substitution}\label{res_lasub}

Recently, a similar valence instability, as found in the system EuPd$_3$B$_x$ \cite{gumeniuk10}, was observed for the compound Eu$_{0.4}$La$_{0.6}$Pd$_3$ where the influence of chemical pressure on the Eu valence state was investigated by susceptibility and XRD measurements \cite{la_pap}. 
The substitution of Eu by La changes the "non-magnetic" EuPd$_3$ into a "magnetic" Eu$_{0.4}$La$_{0.6}$Pd$_3$, comparable to the effect of B insertion in EuPd$_3$B$_x$ for $x\ge 0.2$.
Applying our calculational approach, we find a stable Eu$^{2+}$ state for Eu$_{0.4}$La$_{0.6}$Pd$_3$, in agreement with the reported experiments.
But as La and B are inserted at different crystallographic sites in the systems Eu$_{1-x}$La$_x$Pd$_3$ and EuPd$_3$B$_x$, respectively, the valence instability should be independent from changes of the structure type and the local environment of Eu.
    
\begin{figure}[htb]
\begin{center}\includegraphics[%
  clip,
  width=8cm,
  angle=0]{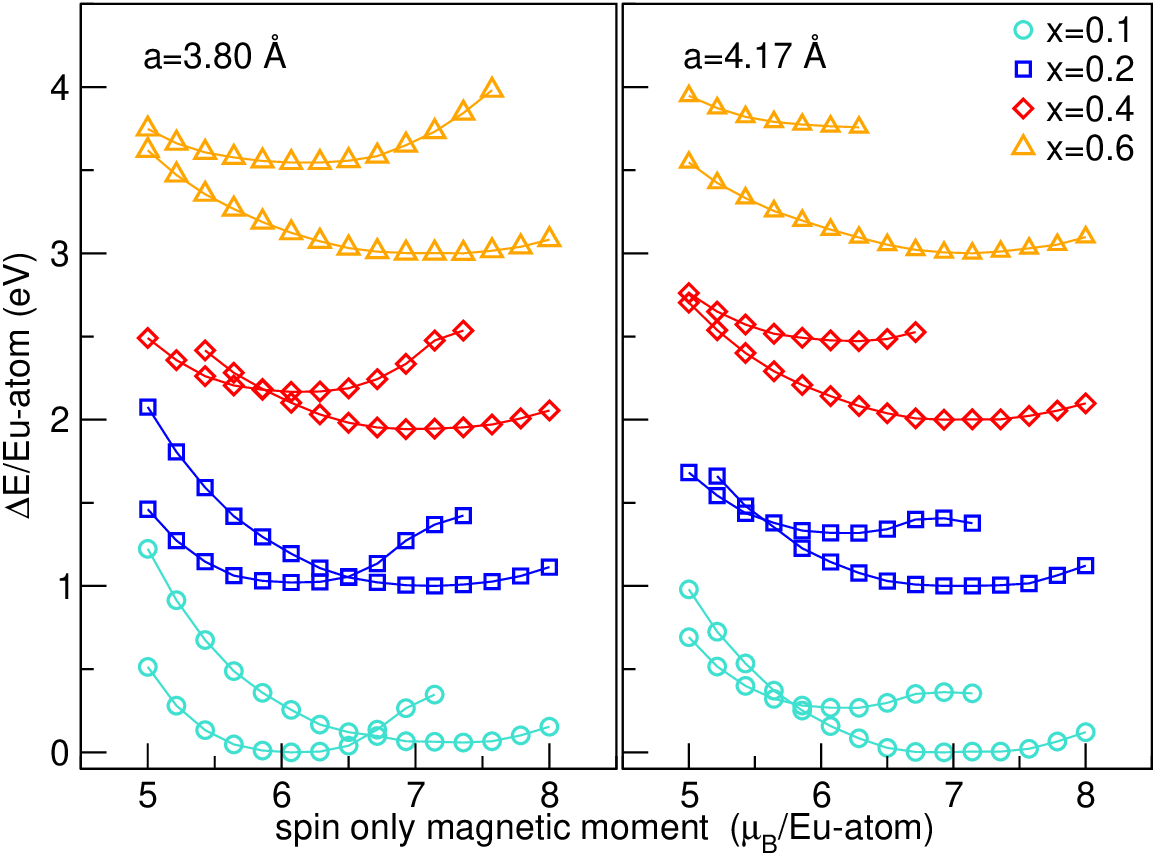}\end{center}
\caption{\label{la_fsm}(color online)
Fixed spin moment calculation for different La concentrations in Eu$_{1-x}$La$_x$Pd$_3$ at two volumes. 
For a small volume ($a=3.80$\,\AA) an increasing La concentration changes the energy balance between the $4f^6$ and $4f^7$ state (left). A large volume 
($a=4.17$\,\AA)
stabilizes the $4f^7$ state (right). }
\end{figure}

For a more detailed analysis we simulated the gradual substitution of Eu by La using the LSDA+$U$+CPA approach, analogous to the case of B substitution.
To study the stability of the valence state, we applied the fixed spin moment method at two different volumes (the experimentally observed and a reduced volume) and varied the La content,
which is equivalent to changing the number of valence electrons in the system. The comparison of the total energies corresponding to an Eu $4f$ occupation close to $6\mu_B/$Eu (for the spin-only contribution) and $7\mu_B/$Eu are depicted in Fig.~\ref{la_fsm}.
At both volumes the La substitution strongly influences the balance between the Eu$^{2+}$ and Eu$^{3+}$ state.
For a small volume ($a=3.80$\AA), the increasing La content shifts the global energy minima from 6 to 7$\mu_B$/Eu, clearly stabilizing the Eu $4f^7$ state for a La concentration of $x\ge 0.4$.
For the experimentally observed volume of Eu$_{0.4}$La$_{0.6}$Pd$_3$ ($a=4.17$\AA) the global energy minimum around 7$\mu_B$/Eu remains stable. Nevertheless, also for the larger volume, the energy difference between the local minima and therefore the Eu $4f^6$ and Eu $4f^7$ state changes significantly depending on the La content. 

\begin{figure}[htb]
\begin{center}\includegraphics[%
  clip,
  width=8cm,
  angle=0]{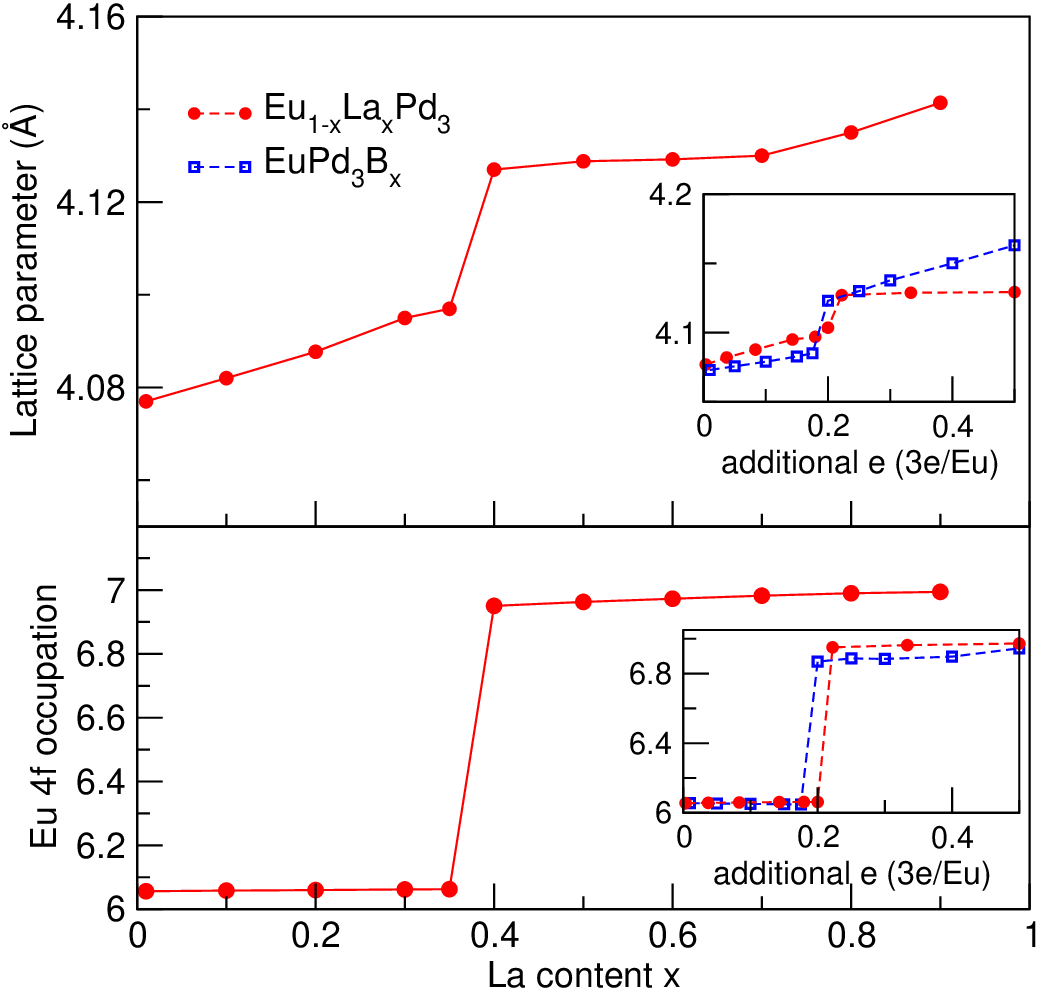}\end{center}
\caption{\label{lll}(color online)
Optimized lattice parameter and Eu $4f$ occupation for different La concentrations in Eu$_{1-x}$La$_x$Pd$_3$. The jump 
in the lattice parameter and Eu $4f$ occupancy coincides. 
Insets: Comparison to the development of EuPd$_3$B$_x$ for different B contents. The data are presented with respect to the same 
number of additional electrons per Eu site in the system.
}
\end{figure}

Furthermore we fully optimized the whole substitution series of Eu$_{1-x}$La$_x$Pd$_3$ compounds, obtaining the volume and Eu $4f$ occupation self-consistently.
Similar to the  observation for the EuPd$_3$B$_x$ series under B insertion, we found a sudden change of the optimized lattice parameters for a critical La concentration $x\approx 0.4$ (compare Fig.~\ref{lll}, top). This discontinuity of the volume is connected to a sudden change in the occupation of the Eu $4f$ majority spin channel (see Fig.~\ref{lll}, bottom).
The comparison of the critical concentration in both systems, EuPd$_3$B$_x$ and Eu$_{1-x}$La$_x$Pd$_3$, reveals the electron count as the key parameter determining the valence state of the systems.
In the case of EuPd$_3$B the system gains 3 valence electrons compared to EuPd$_3$. 
Thus, for the critical B content of $x_c=0.2$, the valence transition sets in at 0.6 electrons per Eu site. Substituting La in Eu$_{1-x}$La$_x$Pd$_3$ not only increases the number of valence electrons but also reduces the number of Eu sites. 
 Based on one additional valence electron per substituted La atom, the critical La concentration between $x=0.35$ and $x=0.4$ (equivalent to 0.35 and 0.4 electrons, respectively) is shared by 0.6 Eu sites, resulting in about 0.54 to 0.67 additional valence electrons per Eu site.
Thus, the critical number of additional electrons per Eu site is essentially the same in both systems (compare Fig.~\ref{lll}, insets), which supports independently the predominant role of the valence electron count for the 
Eu valence change.

\section{summary}\label{sum}

In this joint theoretical and experimental investigation we combined DFT based electronic structure calculations
 with x-ray absorption and x-ray diffraction measurements at high pressures to
elucidate the driving force of the valence transition
in the system EuPd$_3$B$_x$, which stimulated
many studies in the past.
In the series of EuPd$_3$B$_x$  compounds  ($0\le x\le0.53$),
 a  valence transition from Eu$^{3+}$ towards Eu$^{2+}$ into a heterogeneous 
 mixed-valence state was observed
with increasing B content, where the onset of the transition at $x_c=0.2$ yields a 
pronounced lattice anomaly. The strong interplay of the Eu valence state and the crystal structure raised the question 
whether this 
transition is driven by (i) the change in the  volume, (ii) the specific chemical bonding of boron or by (iii) the valence electron count.
Since the substitution of B into the system influences several parameters simultaneously,
in particular the crystal structure and the number of valence electrons,
the underlying mechanism is not obvious. Furthermore, the separation of the different parameters
is impeded as the insertion of an atom usually induces at the same time disorder and changes of the local crystal field due to local distortions.
However, also theory cannot solve the problem unambiguously as a mixed-valence state cannot be simulated 
in a standard DFT approach.
Thus, we applied a combination of different techniques to unravel the contributions of the
different effects on the valence transition. 

To estimate the pure volume effect (i) on the Eu valence state, high-pressure experiments 
for samples with a fixed B content above the critical concentration have been performed.
The results of the XAS measurements yield a sizable influence of pressure on the Eu valence, 
although
the mixed-valence state can be reversed only partially for pressures up to 25\,GPa.
The critical volume for the transition (corresponding to $x=0.2$ at ambient pressure) is already reached 
at about 10\,GPa, but the mean valence of EuPd$_3$B$_{0.48}$ only changes by about 10\%.
This leads to the picture that the volume change (i) is of minor importance for the transition. 
However, the observed valence change of Eu is independently confirmed by XRD experiments under pressure.
 The evaluation of the obtained 
equations of state for two EuPd$_3$B$_x$ ($x=$ 0.32, 0.48)  and two Gd-reference ($x=$ 0.35, 0.44) systems with a stable 
$4f$ configuration resulted in a significantly smaller bulk modulus for the Eu compounds. This softer pressure dependence
of the Eu compounds is a fingerprint of the valence change.
The experimental results conform well with the electronic structure calculations that yield a much stronger influence of the B insertion
than a mere volume expansion of the unit cell.

The relevance of the remaining parameters ((ii) boron chemistry {\it vs.} (iii) valence electron count) could be estimated
by a systematic theoretical study of the series Eu$_{1-x}$La$_x$Pd$_3$, where a similar valence change for x=0.6 was 
reported recently \cite{la_pap}.
The calculations indicate a valence transition in Eu$_{1-x}$La$_x$Pd$_3$ for $x^{\mathrm{La}}_c\ge0.4$ which corresponds 
surprisingly well with the electron count per Eu
for the critical B content $x^{\mathrm{B}}_c=0.2$.  Consequently, this leads to the conclusion
that the valence transition in EuPd$_3$ derived compounds is essentially of electronic origin and ruled by the number 
of additional valence electrons.  
A detailed experimental study of  the Eu$_{1-x}$La$_x$Pd$_3$  system to challenge
the theoretical prediction is underway.

\section{acknowledgement}

We thank for the use of the computational facilities at the 
IFW, Dresden and the access to the ESRF beam lines ID24 (proposals HE-3201 and IH-HC-1243) and ID09. 

%

\bibliographystyle{apsrev4-1}
\bibliography{refs}

\begin{thebibliography}{10}%
\makeatletter
\providecommand \@ifxundefined [1]{%
 \ifx #1\undefined \expandafter \@firstoftwo
 \else \expandafter \@secondoftwo
\fi
}%
\providecommand \@ifnum [1]{%
 \ifnum #1\expandafter \@firstoftwo
 \else \expandafter \@secondoftwo
\fi
}%
\providecommand \enquote [1]{``#1''}%
\providecommand \bibnamefont  [1]{#1}%
\providecommand \bibfnamefont [1]{#1}%
\providecommand \citenamefont [1]{#1}%
\providecommand\href[0]{\@sanitize\@href}%
\providecommand\@href[1]{\endgroup\@@startlink{#1}\endgroup\@@href}%
\providecommand\@@href[1]{#1\@@endlink}%
\providecommand \@sanitize [0]{\begingroup\catcode`\&12\catcode`\#12\relax}%
\@ifxundefined \pdfoutput {\@firstoftwo}{%
 \@ifnum{\z@=\pdfoutput}{\@firstoftwo}{\@secondoftwo}%
}{%
 \providecommand\@@startlink[1]{\leavevmode\special{html:<a href="#1">}}%
 \providecommand\@@endlink[0]{\special{html:</a>}}%
}{%
 \providecommand\@@startlink[1]{%
  \leavevmode
  \pdfstartlink
   attr{/Border[0 0 1 ]/H/I/C[0 1 1]}%
   user{/Subtype/Link/A<</Type/Action/S/URI/URI(#1)>>}%
  \relax
 }%
 \providecommand\@@endlink[0]{\pdfendlink}%
}%
\providecommand \url  [0]{\begingroup\@sanitize \@url }%
\providecommand \@url [1]{\endgroup\@href {#1}{\urlprefix}}%
\providecommand \urlprefix [0]{URL }%
\providecommand \Eprint[0]{\href }%
\@ifxundefined \urlstyle {%
  \providecommand \doi [1]{doi:\discretionary{}{}{}#1}%
}{%
  \providecommand \doi [0]{doi:\discretionary{}{}{}\begingroup
  \urlstyle{rm}\Url }%
}%
\providecommand \doibase [0]{http://dx.doi.org/}%
\providecommand \Doi[1]{\href{\doibase#1}}%
\providecommand \bibAnnote [3]{%
  \BibitemShut{#1}%
  \begin{quotation}\noindent
    \textsc{Key:}\ #2\\\textsc{Annotation:}\ #3%
  \end{quotation}%
}%
\providecommand \bibAnnoteFile [2]{%
  \IfFileExists{#2}{\bibAnnote {#1} {#2} {\input{#2}}}{}%
}%
\providecommand \typeout [0]{\immediate \write \m@ne }%
\providecommand \selectlanguage [0]{\@gobble}%
\providecommand \bibinfo [0]{\@secondoftwo}%
\providecommand \bibfield [0]{\@secondoftwo}%
\providecommand \translation [1]{[#1]}%
\providecommand \BibitemOpen[0]{}%
\providecommand \bibitemStop [0]{}%
\providecommand \bibitemNoStop [0]{.\EOS\space}%
\providecommand \EOS [0]{\spacefactor3000\relax}%
\providecommand \BibitemShut [1]{\csname bibitem#1\endcsname}%
\bibitem{Bauminger1974}%
  \BibitemOpen
  \bibfield{author}{%
  \bibinfo {author} {\bibfnamefont{E.~R.}\ \bibnamefont{Bauminger}}, \bibinfo
  {author} {\bibfnamefont{I.}~\bibnamefont{Felner}}, \bibinfo {author}
  {\bibfnamefont{D.}~\bibnamefont{Levron}}, \bibinfo {author}
  {\bibfnamefont{I.}~\bibnamefont{Nowik}},\ and\ \bibinfo {author}
  {\bibfnamefont{S.}~\bibnamefont{Ofer}},\ }%
  \bibfield{journal}{%
  \bibinfo {journal} {Phys. Rev. Lett.}\ }%
  \textbf{\bibinfo {volume} {33}},\ \bibinfo {pages} {890} (\bibinfo {year}
  {1974})%
  \bibAnnoteFile{NoStop}{Bauminger1974}%
\bibitem{Felner1977}%
  \BibitemOpen
  \bibfield{author}{%
  \bibinfo {author} {\bibfnamefont{I.}~\bibnamefont{Felner}}\ and\ \bibinfo
  {author} {\bibfnamefont{I.}~\bibnamefont{Nowik}},\ }%
  \bibfield{journal}{%
  \bibinfo {journal} {J. Phys. Chem. Solids}\ }%
  \textbf{\bibinfo {volume} {39}},\ \bibinfo {pages} {763} (\bibinfo {year}
  {1977})%
  \bibAnnoteFile{NoStop}{Felner1977}%
\bibitem{Sampathkumaran1981}%
  \BibitemOpen
  \bibfield{author}{%
  \bibinfo {author} {\bibfnamefont{E.~V.}\ \bibnamefont{Sampathkumaran}},
  \bibinfo {author} {\bibfnamefont{L.~C.}\ \bibnamefont{Gupta}}, \bibinfo
  {author} {\bibfnamefont{R.}~\bibnamefont{Vijayaraghavan}}, \bibinfo {author}
  {\bibfnamefont{K.~V.}\ \bibnamefont{Gopalakrishnan}}, \bibinfo {author}
  {\bibfnamefont{R.}~\bibnamefont{Pillay}},\ and\ \bibinfo {author}
  {\bibfnamefont{H.~G.}\ \bibnamefont{Devare}},\ }%
  \bibfield{journal}{%
  \bibinfo {journal} {J. Phys.}\ }%
  \textbf{\bibinfo {volume} {C14}},\ \bibinfo {pages} {L237} (\bibinfo {year}
  {1981})%
  \bibAnnoteFile{NoStop}{Sampathkumaran1981}%
\bibitem{Perscheid1985}%
  \BibitemOpen
  \bibfield{author}{%
  \bibinfo {author} {\bibfnamefont{B.}~\bibnamefont{Perscheid}}, \bibinfo
  {author} {\bibfnamefont{E.~V.}\ \bibnamefont{Sampathkumaran}},\ and\ \bibinfo
  {author} {\bibfnamefont{G.}~\bibnamefont{Kaindl}},\ }%
  \bibfield{journal}{%
  \bibinfo {journal} {J. Mag. Magn. Mater.}\ }%
  \textbf{\bibinfo {volume} {47}},\ \bibinfo {pages} {410} (\bibinfo {year}
  {1985})%
  \bibAnnoteFile{NoStop}{Perscheid1985}%
\bibitem{Nagarajan1981}%
  \BibitemOpen
  \bibfield{author}{%
  \bibinfo {author} {\bibfnamefont{R.}~\bibnamefont{Nagarajan}}, \bibinfo
  {author} {\bibfnamefont{E.~V.}\ \bibnamefont{Sampathkumaran}}, \bibinfo
  {author} {\bibfnamefont{L.~C.}\ \bibnamefont{Gupta}}, \bibinfo {author}
  {\bibfnamefont{R.}~\bibnamefont{Vijayaraghavan}}, \bibinfo {author}
  {\bibfnamefont{V.}~\bibnamefont{Prabhawalkar}}, \bibinfo {author}
  {\bibfnamefont{B.}~\bibnamefont{Haktdarshen}},\ and\ \bibinfo {author}
  {\bibfnamefont{B.~D.}\ \bibnamefont{Padalia}},\ }%
  \bibfield{journal}{%
  \bibinfo {journal} {Physics Lett.}\ }%
  \textbf{\bibinfo {volume} {84A}},\ \bibinfo {pages} {275} (\bibinfo {year}
  {1981})%
  \bibAnnoteFile{NoStop}{Nagarajan1981}%
\bibitem{Wortmann1991}%
  \BibitemOpen
  \bibfield{author}{%
  \bibinfo {author} {\bibfnamefont{G.}~\bibnamefont{Wortmann}}, \bibinfo
  {author} {\bibfnamefont{I.}~\bibnamefont{Nowik}}, \bibinfo {author}
  {\bibfnamefont{B.}~\bibnamefont{Perscheid}}, \bibinfo {author}
  {\bibfnamefont{G.}~\bibnamefont{Kaindl}},\ and\ \bibinfo {author}
  {\bibfnamefont{I.}~\bibnamefont{Felner}},\ }%
  \bibfield{journal}{%
  \bibinfo {journal} {Phys. Rev. B}\ }%
  \textbf{\bibinfo {volume} {43}},\ \bibinfo {pages} {5261} (\bibinfo {year}
  {1991})%
  \bibAnnoteFile{NoStop}{Wortmann1991}%
\bibitem{Wada1999}%
  \BibitemOpen
  \bibfield{author}{%
  \bibinfo {author} {\bibfnamefont{H.}~\bibnamefont{Wada}}, \bibinfo {author}
  {\bibfnamefont{M.~F.}\ \bibnamefont{Hundley}}, \bibinfo {author}
  {\bibfnamefont{R.}~\bibnamefont{Movshovich}},\ and\ \bibinfo {author}
  {\bibfnamefont{J.~D.}\ \bibnamefont{Thompson}},\ }%
  \bibfield{journal}{%
  \bibinfo {journal} {Phys. Rev. B}\ }%
  \textbf{\bibinfo {volume} {59}},\ \bibinfo {pages} {1141} (\bibinfo {year}
  {1999})%
  \bibAnnoteFile{NoStop}{Wada1999}%
\bibitem{Ni2001}%
  \BibitemOpen
  \bibfield{author}{%
  \bibinfo {author} {\bibfnamefont{B.}~\bibnamefont{Ni}}, \bibinfo {author}
  {\bibfnamefont{M.~M.}\ \bibnamefont{Abd-Elmeguid}}, \bibinfo {author}
  {\bibfnamefont{H.}~\bibnamefont{Micklitz}}, \bibinfo {author}
  {\bibfnamefont{J.~P.}\ \bibnamefont{Sanchez}}, \bibinfo {author}
  {\bibfnamefont{P.}~\bibnamefont{Vulliet}},\ and\ \bibinfo {author}
  {\bibfnamefont{D.}~\bibnamefont{Johrendt}},\ }%
  \bibfield{journal}{%
  \bibinfo {journal} {Phys. Rev. B}\ }%
  \textbf{\bibinfo {volume} {63}},\ \bibinfo {pages} {100102 (R)} (\bibinfo
  {year} {2001})%
  \bibAnnoteFile{NoStop}{Ni2001}%
\bibitem{Hossain2004}%
  \BibitemOpen
  \bibfield{author}{%
  \bibinfo {author} {\bibfnamefont{Z.}~\bibnamefont{Hossain}}, \bibinfo
  {author} {\bibfnamefont{C.}~\bibnamefont{Geibel}}, \bibinfo {author}
  {\bibfnamefont{N.}~\bibnamefont{Senthilkumaran}}, \bibinfo {author}
  {\bibfnamefont{M.}~\bibnamefont{Deppe}}, \bibinfo {author}
  {\bibfnamefont{M.}~\bibnamefont{Baenitz}}, \bibinfo {author}
  {\bibfnamefont{F.}~\bibnamefont{Schiller}},\ and\ \bibinfo {author}
  {\bibfnamefont{S.~L.}\ \bibnamefont{Molodtsov}},\ }%
  \bibfield{journal}{%
  \bibinfo {journal} {Phys. Rev. B}\ }%
  \textbf{\bibinfo {volume} {69}},\ \bibinfo {pages} {014422} (\bibinfo {year}
  {2004})%
  \bibAnnoteFile{NoStop}{Hossain2004}%
\bibitem{Michels1995}%
  \BibitemOpen
  \bibfield{author}{%
  \bibinfo {author} {\bibfnamefont{G.}~\bibnamefont{Michels}}, \bibinfo
  {author} {\bibfnamefont{C.}~\bibnamefont{Huhnt}}, \bibinfo {author}
  {\bibfnamefont{W.}~\bibnamefont{Scharbrodt}}, \bibinfo {author}
  {\bibfnamefont{W.}~\bibnamefont{Schlabitz}}, \bibinfo {author}
  {\bibfnamefont{E.}~\bibnamefont{Holland-Moritz}}, \bibinfo {author}
  {\bibfnamefont{M.~M.}\ \bibnamefont{Abd-Elmeguid}}, \bibinfo {author}
  {\bibfnamefont{H.}~\bibnamefont{Micklitz}}, \bibinfo {author}
  {\bibfnamefont{D.}~\bibnamefont{Johrendt}}, \bibinfo {author}
  {\bibfnamefont{V.}~\bibnamefont{Keimes}},\ and\ \bibinfo {author}
  {\bibfnamefont{A.}~\bibnamefont{Mewis}},\ }%
  \bibfield{journal}{%
  \bibinfo {journal} {Z. Phys. B}\ }%
  \textbf{\bibinfo {volume} {98}},\ \bibinfo {pages} {75} (\bibinfo {year}
  {1995})%
  \bibAnnoteFile{NoStop}{Michels1995}%
\bibitem{Poettgen2000}%
  \BibitemOpen
  \bibfield{author}{%
  \bibinfo {author} {\bibfnamefont{R.}~\bibnamefont{P{\"o}ttgen}}\ and\
  \bibinfo {author} {\bibfnamefont{D.}~\bibnamefont{Johrendt}},\ }%
  \bibfield{journal}{%
  \bibinfo {journal} {Chem. Mater.}\ }%
  \textbf{\bibinfo {volume} {12}},\ \bibinfo {pages} {875} (\bibinfo {year}
  {2000})%
  \bibAnnoteFile{NoStop}{Poettgen2000}%
\bibitem{Klauss1997}%
  \BibitemOpen
  \bibfield{author}{%
  \bibinfo {author} {\bibfnamefont{H.~H.}\ \bibnamefont{Klauss}}, \bibinfo
  {author} {\bibfnamefont{M.}~\bibnamefont{Hillberg}}, \bibinfo {author}
  {\bibfnamefont{W.}~\bibnamefont{Wagner}}, \bibinfo {author}
  {\bibfnamefont{M.~A.~C.}\ \bibnamefont{de~Melo}}, \bibinfo {author}
  {\bibfnamefont{F.~J.}\ \bibnamefont{Litterst}}, \bibinfo {author}
  {\bibfnamefont{E.}~\bibnamefont{Schreier}}, \bibinfo {author}
  {\bibfnamefont{W.}~\bibnamefont{Schlabitz}},\ and\ \bibinfo {author}
  {\bibfnamefont{G.}~\bibnamefont{Michels}},\ }%
  \bibfield{journal}{%
  \bibinfo {journal} {Hyperfine Interactions}\ }%
  \textbf{\bibinfo {volume} {104}},\ \bibinfo {pages} {171} (\bibinfo {year}
  {1997})%
  \bibAnnoteFile{NoStop}{Klauss1997}%
\bibitem{Ksenofontov2006}%
  \BibitemOpen
  \bibfield{author}{%
  \bibinfo {author} {\bibfnamefont{V.}~\bibnamefont{Ksenofontov}}, \bibinfo
  {author} {\bibfnamefont{H.~C.}\ \bibnamefont{Kandpal}}, \bibinfo {author}
  {\bibfnamefont{J.}~\bibnamefont{Ensling}}, \bibinfo {author}
  {\bibfnamefont{M.}~\bibnamefont{Waldeck}}, \bibinfo {author}
  {\bibfnamefont{D.}~\bibnamefont{Johrendt}}, \bibinfo {author}
  {\bibfnamefont{A.}~\bibnamefont{Mewis}}, \bibinfo {author}
  {\bibfnamefont{P.~G.}\ \bibnamefont{G{\"u}tlich}},\ and\ \bibinfo {author}
  {\bibfnamefont{C.}~\bibnamefont{Felser}},\ }%
  \bibfield{journal}{%
  \bibinfo {journal} {Europhys. Lett.}\ }%
  \textbf{\bibinfo {volume} {74}},\ \bibinfo {pages} {672} (\bibinfo {year}
  {2006})%
  \bibAnnoteFile{NoStop}{Ksenofontov2006}%
\bibitem{Gschneidner1969}%
  \BibitemOpen
  \bibfield{author}{%
  \bibinfo {author} {\bibfnamefont{K.~A.}\ \bibnamefont{Gschneidner}},\ }%
  \bibfield{journal}{%
  \bibinfo {journal} {J. Less Comm. Met.}\ }%
  \textbf{\bibinfo {volume} {17}},\ \bibinfo {pages} {13} (\bibinfo {year}
  {1969})%
  \bibAnnoteFile{NoStop}{Gschneidner1969}%
\bibitem{Buschow1979}%
  \BibitemOpen
  \bibfield{author}{%
  \bibinfo {author} {\bibfnamefont{K.~H.~J.}\ \bibnamefont{Buschow}},\ }%
  \bibfield{journal}{%
  \bibinfo {journal} {Rep. Prog. Physics}\ }%
  \textbf{\bibinfo {volume} {42}},\ \bibinfo {pages} {1373} (\bibinfo {year}
  {1979})%
  \bibAnnoteFile{NoStop}{Buschow1979}%
\bibitem{M76}%
  \BibitemOpen
  \bibfield{author}{%
  \bibinfo {author} {\bibfnamefont{A.~R.}\ \bibnamefont{Miedema}},\ }%
  \bibfield{journal}{%
  \bibinfo {journal} {J. Less Com. Met.}\ }%
  \textbf{\bibinfo {volume} {46}},\ \bibinfo {pages} {167} (\bibinfo {year}
  {1976})%
  \bibAnnoteFile{NoStop}{M76}%
\bibitem{B79}%
  \BibitemOpen
  \bibfield{author}{%
  \bibinfo {author} {\bibfnamefont{F.~R.}\ \bibnamefont{de~Boer}}, \bibinfo
  {author} {\bibfnamefont{W.~H.}\ \bibnamefont{Dijkman}}, \bibinfo {author}
  {\bibfnamefont{W.~C.~M.}\ \bibnamefont{Mattens}},\ and\ \bibinfo {author}
  {\bibfnamefont{A.~R.}\ \bibnamefont{Miedema}},\ }%
  \bibfield{journal}{%
  \bibinfo {journal} {J. Less Common Met.}\ }%
  \textbf{\bibinfo {volume} {64}},\ \bibinfo {pages} {241} (\bibinfo {year}
  {1979})%
  \bibAnnoteFile{NoStop}{B79}%
\bibitem{M83}%
  \BibitemOpen
  \bibfield{author}{%
  \bibinfo {author} {\bibfnamefont{W.~C.~M.}\ \bibnamefont{Mattens}}, \bibinfo
  {author} {\bibfnamefont{F.~R.}\ \bibnamefont{de~Boer}}, \bibinfo {author}
  {\bibfnamefont{A.~K.}\ \bibnamefont{Nissen}},\ and\ \bibinfo {author}
  {\bibfnamefont{A.~R.}\ \bibnamefont{Miedema}},\ }%
  \bibfield{journal}{%
  \bibinfo {journal} {J. Magn. Magn. Mater.}\ }%
  \textbf{\bibinfo {volume} {31-34}},\ \bibinfo {pages} {451} (\bibinfo {year}
  {1983})%
  \bibAnnoteFile{NoStop}{M83}%
\bibitem{L81}%
  \BibitemOpen
  \bibfield{author}{%
  \bibinfo {author} {\bibfnamefont{J.~M.}\ \bibnamefont{Lawrence}}, \bibinfo
  {author} {\bibfnamefont{P.~S.}\ \bibnamefont{Riseborough}},\ and\ \bibinfo
  {author} {\bibfnamefont{R.~D.}\ \bibnamefont{Parks}},\ }%
  \bibfield{journal}{%
  \bibinfo {journal} {Rep. Prog. Physics}\ }%
  \textbf{\bibinfo {volume} {44}},\ \bibinfo {pages} {1} (\bibinfo {year}
  {1981})%
  \bibAnnoteFile{NoStop}{L81}%
\bibitem{gumeniuk10}%
  \BibitemOpen
  \bibfield{author}{%
  \bibinfo {author} {\bibfnamefont{R.}~\bibnamefont{Gumeniuk}}, \bibinfo
  {author} {\bibfnamefont{M.}~\bibnamefont{Schmitt}}, \bibinfo {author}
  {\bibfnamefont{C.}~\bibnamefont{Loison}}, \bibinfo {author}
  {\bibfnamefont{W.}~\bibnamefont{Carrillo-Cabrera}}, \bibinfo {author}
  {\bibfnamefont{U.}~\bibnamefont{Burkhardt}}, \bibinfo {author}
  {\bibfnamefont{G.}~\bibnamefont{Auffermann}}, \bibinfo {author}
  {\bibfnamefont{M.}~\bibnamefont{Schmidt}}, \bibinfo {author}
  {\bibfnamefont{W.}~\bibnamefont{Schnelle}}, \bibinfo {author}
  {\bibfnamefont{C.}~\bibnamefont{Geibel}}, \bibinfo {author}
  {\bibfnamefont{A.}~\bibnamefont{Leithe-Jasper}},\ and\ \bibinfo {author}
  {\bibfnamefont{H.}~\bibnamefont{Rosner}},\ }%
  \bibfield{journal}{%
  \bibinfo {journal} {Phys. Rev. B}\ }%
  \textbf{\bibinfo {volume} {82}},\ \bibinfo {pages} {235113} (\bibinfo {year}
  {2010})%
  \bibAnnoteFile{NoStop}{gumeniuk10}%
\bibitem{CL07}%
  \BibitemOpen
  \bibfield{author}{%
  \bibinfo {author} {\bibfnamefont{C.}~\bibnamefont{Loison}}, \bibinfo {author}
  {\bibfnamefont{A.}~\bibnamefont{Leithe-Jasper}},\ and\ \bibinfo {author}
  {\bibfnamefont{H.}~\bibnamefont{Rosner}},\ }%
  \bibfield{journal}{%
  \bibinfo {journal} {Phys. Rev. B}\ }%
  \textbf{\bibinfo {volume} {75}},\ \bibinfo {pages} {205135} (\bibinfo {year}
  {2007})%
  \bibAnnoteFile{NoStop}{CL07}%
\bibitem{la_pap}%
  \BibitemOpen
  \bibfield{author}{%
  \bibinfo {author} {\bibfnamefont{A.}~\bibnamefont{Pandey}}, \bibinfo {author}
  {\bibfnamefont{C.}~\bibnamefont{Mazumdar}},\ and\ \bibinfo {author}
  {\bibfnamefont{R.}~\bibnamefont{Ranganathan}},\ }%
  \bibfield{journal}{%
  \bibinfo {journal} {J. Phys.: Condens. Matter}\ }%
  \textbf{\bibinfo {volume} {21}},\ \bibinfo {pages} {216002} (\bibinfo {year}
  {2009})%
  \bibAnnoteFile{NoStop}{la_pap}%
\bibitem{fplo1}%
  \BibitemOpen
  \bibfield{author}{%
  \bibinfo {author} {\bibfnamefont{K.}~\bibnamefont{Koepernik}}\ and\ \bibinfo
  {author} {\bibfnamefont{H.}~\bibnamefont{Eschrig}},\ }%
  \bibfield{journal}{%
  \bibinfo {journal} {Phys. Rev. B}\ }%
  \textbf{\bibinfo {volume} {59}},\ \bibinfo {pages} {1743} (\bibinfo {year}
  {1999})%
  \bibAnnoteFile{NoStop}{fplo1}%
\bibitem{PW92}%
  \BibitemOpen
  \bibfield{author}{%
  \bibinfo {author} {\bibfnamefont{J.~P.}\ \bibnamefont{Perdew}}\ and\ \bibinfo
  {author} {\bibfnamefont{Y.}~\bibnamefont{Wang}},\ }%
  \bibfield{journal}{%
  \bibinfo {journal} {Phys. Rev. B}\ }%
  \textbf{\bibinfo {volume} {45}},\ \bibinfo {pages} {13244} (\bibinfo {year}
  {1992})%
  \bibAnnoteFile{NoStop}{PW92}%
\bibitem{fplo2}%
  \BibitemOpen
  \bibfield{author}{%
  \bibinfo {author} {\bibfnamefont{H.}~\bibnamefont{Eschrig}}, \bibinfo
  {author} {\bibfnamefont{K.}~\bibnamefont{Koepernik}},\ and\ \bibinfo {author}
  {\bibfnamefont{I.}~\bibnamefont{Chaplygin}},\ }%
  \bibfield{journal}{%
  \bibinfo {journal} {J. Solid States Chemistry}\ }%
  \textbf{\bibinfo {volume} {176}},\ \bibinfo {pages} {482} (\bibinfo {year}
  {2003})%
  \bibAnnoteFile{NoStop}{fplo2}%
\bibitem{koep97}%
  \BibitemOpen
  \bibfield{author}{%
  \bibinfo {author} {\bibfnamefont{K.}~\bibnamefont{Koepernik}}, \bibinfo
  {author} {\bibfnamefont{B.}~\bibnamefont{Velicky}}, \bibinfo {author}
  {\bibfnamefont{R.}~\bibnamefont{Hayn}},\ and\ \bibinfo {author}
  {\bibfnamefont{H.}~\bibnamefont{Eschrig}},\ }%
  \bibfield{journal}{%
  \bibinfo {journal} {Phys. Rev. B}\ }%
  \textbf{\bibinfo {volume} {55}},\ \bibinfo {pages} {5717} (\bibinfo {year}
  {1997})%
  \bibAnnoteFile{NoStop}{koep97}%
\bibitem{BEB}%
  \BibitemOpen
  \bibfield{author}{%
  \bibinfo {author} {\bibfnamefont{J.~A.}\ \bibnamefont{Blackman}}, \bibinfo
  {author} {\bibfnamefont{D.~M.}\ \bibnamefont{Esterling}},\ and\ \bibinfo
  {author} {\bibfnamefont{N.~F.}\ \bibnamefont{Berk}},\ }%
  \bibfield{journal}{%
  \bibinfo {journal} {Phys. Rev. B}\ }%
  \textbf{\bibinfo {volume} {4}},\ \bibinfo {pages} {2412} (\bibinfo {year}
  {1971})%
  \bibAnnoteFile{NoStop}{BEB}%
\bibitem{forman72}%
  \BibitemOpen
  \bibfield{author}{%
  \bibinfo {author} {\bibfnamefont{R.~A.}\ \bibnamefont{Forman}}, \bibinfo
  {author} {\bibfnamefont{G.~J.}\ \bibnamefont{Piermarini}}, \bibinfo {author}
  {\bibfnamefont{J.~D.}\ \bibnamefont{Barnett}},\ and\ \bibinfo {author}
  {\bibfnamefont{S.}~\bibnamefont{Block}},\ }%
  \bibfield{journal}{%
  \bibinfo {journal} {Science}\ }%
  \textbf{\bibinfo {volume} {176}},\ \bibinfo {pages} {284} (\bibinfo {year}
  {1972})%
  \bibAnnoteFile{NoStop}{forman72}%
\bibitem{fit2D}%
  \BibitemOpen
  \bibfield{author}{%
  \bibinfo {author} {\bibfnamefont{A.~P.}\ \bibnamefont{Hammersley}}, \bibinfo
  {author} {\bibfnamefont{S.~O.}\ \bibnamefont{Svensson}}, \bibinfo {author}
  {\bibfnamefont{M.}~\bibnamefont{Hanßand}}, \bibinfo {author}
  {\bibfnamefont{A.}~\bibnamefont{Fitch}},\ and\ \bibinfo {author}
  {\bibfnamefont{D.}~\bibnamefont{Haeusermann}},\ }%
  \bibfield{journal}{%
  \bibinfo {journal} {High Press. Res.}\ }%
  \textbf{\bibinfo {volume} {14}},\ \bibinfo {pages} {235} (\bibinfo {year}
  {1996})%
  \bibAnnoteFile{NoStop}{fit2D}%
\bibitem{fullprof}%
  \BibitemOpen
  \bibfield{author}{%
  \bibinfo {author} {\bibfnamefont{J.}~\bibnamefont{Rodriguez-Carvajal}},\ }%
  \bibfield{journal}{%
  \bibinfo {journal} {Physica B}\ }%
  \textbf{\bibinfo {volume} {192}},\ \bibinfo {pages} {55} (\bibinfo {year}
  {1993})%
  \bibAnnoteFile{NoStop}{fullprof}%
\bibitem{Howard2004}%
  \BibitemOpen
  \bibfield{author}{%
  \bibinfo {author} {\bibfnamefont{C.~J.}\ \bibnamefont{Howard}}\ and\ \bibinfo
  {author} {\bibfnamefont{H.~T.}\ \bibnamefont{Stokes}},\ }%
  \bibfield{journal}{%
  \bibinfo {journal} {Acta Cryst.}\ }%
  \textbf{\bibinfo {volume} {A61}},\ \bibinfo {pages} {93} (\bibinfo {year}
  {2005})%
  \bibAnnoteFile{NoStop}{Howard2004}%
\bibitem{Aleksandrov1976}%
  \BibitemOpen
  \bibfield{author}{%
  \bibinfo {author} {\bibfnamefont{K.~S.}\ \bibnamefont{Aleksandrov}},\ }%
  \bibfield{journal}{%
  \bibinfo {journal} {Sov. Phys. Crystallogr.}\ }%
  \textbf{\bibinfo {volume} {21}},\ \bibinfo {pages} {133} (\bibinfo {year}
  {1976})%
  \bibAnnoteFile{NoStop}{Aleksandrov1976}%
\bibitem{note4}%
  \BibitemOpen
  \bibinfo {journal} {See supplementary material document no. XXX.}%
  \bibAnnoteFile{Stop}{note4}%
\bibitem{MizTsu2007}%
  \BibitemOpen
\bibfield{journal}{%
    }%
  \bibfield{author}{%
  \bibinfo {author} {\bibfnamefont{M.}~\bibnamefont{Mizumaki}}, \bibinfo
  {author} {\bibfnamefont{S.}~\bibnamefont{Tsutsui}}, \bibinfo {author}
  {\bibfnamefont{H.}~\bibnamefont{Tanida}}, \bibinfo {author}
  {\bibfnamefont{T.}~\bibnamefont{Uruga}}, \bibinfo {author}
  {\bibfnamefont{D.}~\bibnamefont{Kikuchi}}, \bibinfo {author}
  {\bibfnamefont{H.}~\bibnamefont{Sugawara}},\ and\ \bibinfo {author}
  {\bibfnamefont{H.}~\bibnamefont{Sato}},\ }%
  \bibfield{journal}{%
  \bibinfo {journal} {J. Phys. Soc. Jpn.}\ }%
  \textbf{\bibinfo {volume} {76}},\ \bibinfo {pages} {053706} (\bibinfo {year}
  {2007})%
  \bibAnnoteFile{NoStop}{MizTsu2007}%
\bibitem{note1}%
  \BibitemOpen
  \bibinfo {journal} {We change the unit cell volume corresponding to a change
  in the lattice parameters from 3.80\AA\ to 4.16\AA. This volume range
  corresponds to a change in the lattice parameter from about 0.3\AA\ smaller
  than the optimized lattice parameter for the B free compound EuPd$_3$ to the
  optimized volume for EuPd$_3$B$_{0.5}$ with nearly maximal B concentration.}%
  \bibAnnoteFile{Stop}{note1}%
\bibitem{supp}%
  \BibitemOpen
\bibfield{journal}{%
    }%
  \bibinfo {journal} {See supplementary material for a direct comparison of the
  resulting fixed spin moment calculation using VCA and CPA. Document no.
  XXX.}%
  \bibAnnoteFile{Stop}{supp}%
\bibitem{cianchi91}%
  \BibitemOpen
\bibfield{journal}{%
    }%
  \bibfield{author}{%
  \bibinfo {author} {\bibfnamefont{L.}~\bibnamefont{Cianchi}}, \bibinfo
  {author} {\bibfnamefont{S.}~\bibnamefont{de~Geunera}}, \bibinfo {author}
  {\bibfnamefont{F.}~\bibnamefont{Gulisano}}, \bibinfo {author}
  {\bibfnamefont{M.}~\bibnamefont{Menchini}},\ and\ \bibinfo {author}
  {\bibfnamefont{G.}~\bibnamefont{Spina}},\ }%
  \bibfield{journal}{%
  \bibinfo {journal} {J. Phys.: Condens. Matter}\ }%
  \textbf{\bibinfo {volume} {3}},\ \bibinfo {pages} {781} (\bibinfo {year}
  {1991})%
  \bibAnnoteFile{NoStop}{cianchi91}%
\bibitem{note3}%
  \BibitemOpen
  \bibinfo {journal} {Volume difference from 4\% from theory and 6\% from
  XRD.}%
  \bibAnnoteFile{Stop}{note3}%
\bibitem{note2}%
  \BibitemOpen
\bibfield{journal}{%
    }%
  \bibinfo {journal} {The larger error for the GdPd$_3$B$_x$ compounds results
  from a slightly increase of the peak widths as discussion in
  Sec.~\ref{method} and shown in the supplementary material (Document no.
  XXX.).}%
  \bibAnnoteFile{Stop}{note2}%
\end{thebibliography}%

\end{document}